\documentclass[twocolumn]{aastex62}
\graphicspath{{./}{figures/}}
\usepackage{gensymb}

\newcommand{\kms}{km~s$^{-1}$}

\newcommand{\kmsMpc}{km~s$^{-1}$~Mpc$^{-1}$}

\begin{document}

\title{The Distance and Motion of the Maffei Group}
\shorttitle{The Distance and Motion of the Maffei Group}
\author{Gagandeep S. Anand}
\affil{Institute for Astronomy, University of Hawaii, 2680 Woodlawn Drive, Honolulu, HI 96822, USA}
\author{R. Brent Tully}
\affil{Institute for Astronomy, University of Hawaii, 2680 Woodlawn Drive, Honolulu, HI 96822, USA}
\author{Luca Rizzi}
\affil{W. M. Keck Observatory, 65-1120 Mamalahoa Hwy., Kamuela, HI 96743, USA}
\author{Igor D. Karachentsev}
\affil{Special Astrophysical Observatory, Nizhniy Arkhyz, Karachai-Cherkessia 369167, Russia}

\begin{abstract}
It has recently been suggested that the nearby galaxies Maffei 1 and 2 are further in distance than previously thought, such that they no longer are members of the same galaxy group as IC~342. We reanalyze near-infrared photometry from the Hubble Space Telescope, and find a distance to Maffei~2 of $5.73\pm0.40$~Mpc.  With this distance, the Maffei Group lies 2.5~Mpc behind the IC~342 Group and has a peculiar velocity toward the Local Group of $-128\pm33$~\kms. The negative peculiar velocities of both of these distinct galaxy groups are likely the manifestation of void expansion from the direction of Perseus-Pisces.

\end{abstract}
\keywords{}

\section{Introduction}

Our collaboration \citep{wu2014} determined an averaged distance to the galaxies Maffei 1 \& 2 of 3.4 $\pm$ 0.2 Mpc from Tip of the Red Giant Branch (TRGB) measurements based on near infrared photometry of Hubble Space Telescope (HST) images.  At this distance, the Maffei Group along with the galaxies around IC~342 would be the nearest substantial concentration of galaxies to our Local Group. Subsequently, \cite{tikhonov} have re-evaluated the same HST images and determined an averaged distance of 6.7 $\pm$ 0.5 Mpc for the Maffei pair, twice the Wu et al. value. At issue is potential confusion between the brightest stars of the Red Giant Branch (RGB) and the onset of Asymptotic Giant Branch (AGB) stars, a concern pointed out in other cases \citep{Aloisi2007,antennaeNEW, comap}.  

Here, we have re-analyzed the same HST images.  Our new results confirm the greater distances found by \cite{tikhonov}, although there are important differences in the details.  The larger distance has interesting implications regarding the peculiar velocity of the Maffei Group.  There is agreement that IC~342 is much closer so the Maffei and IC~342 groups are quite distinct.

\begin{figure*}
\gridline{\fig{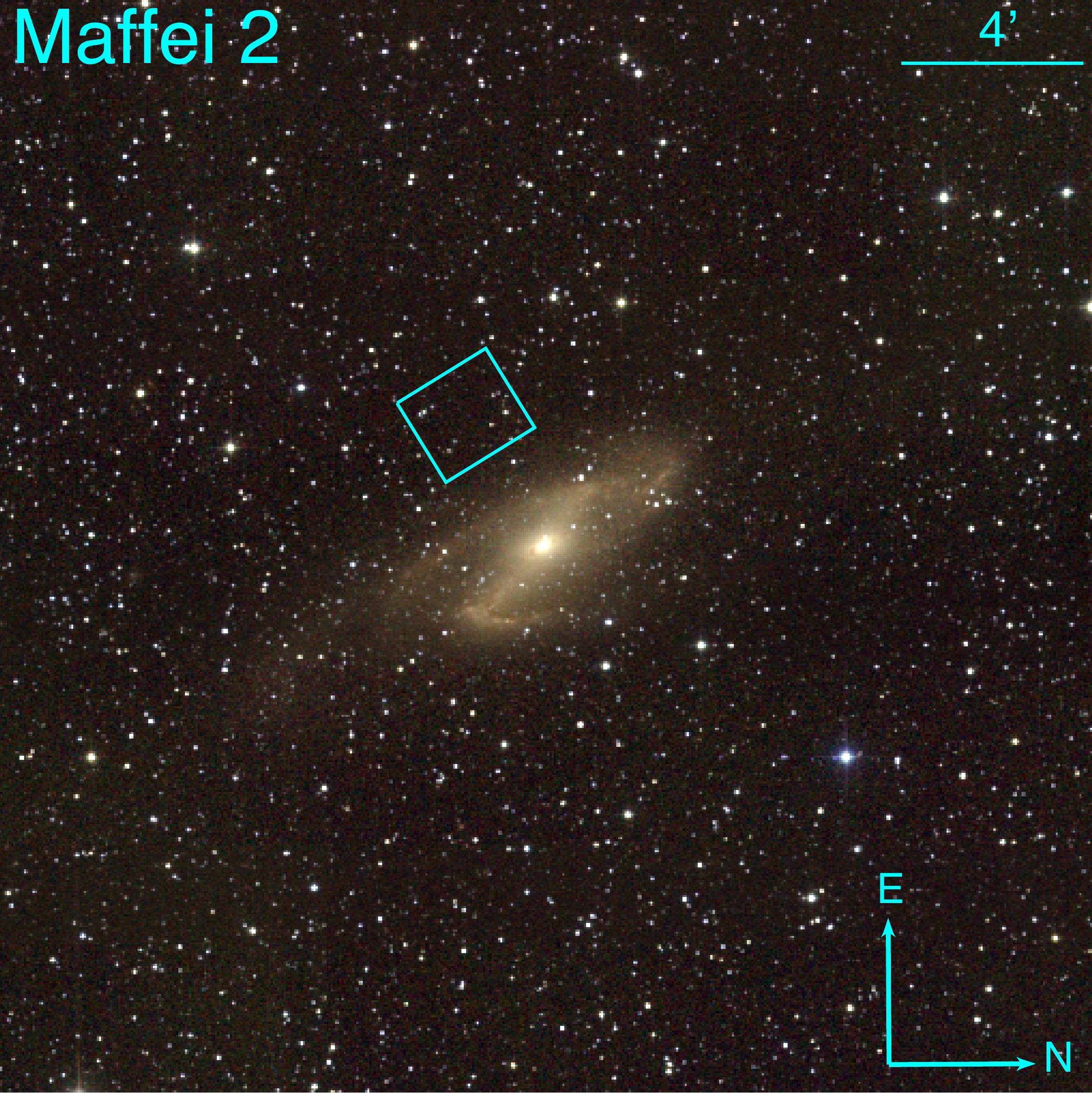}{0.45\textwidth}{(a)}
          \fig{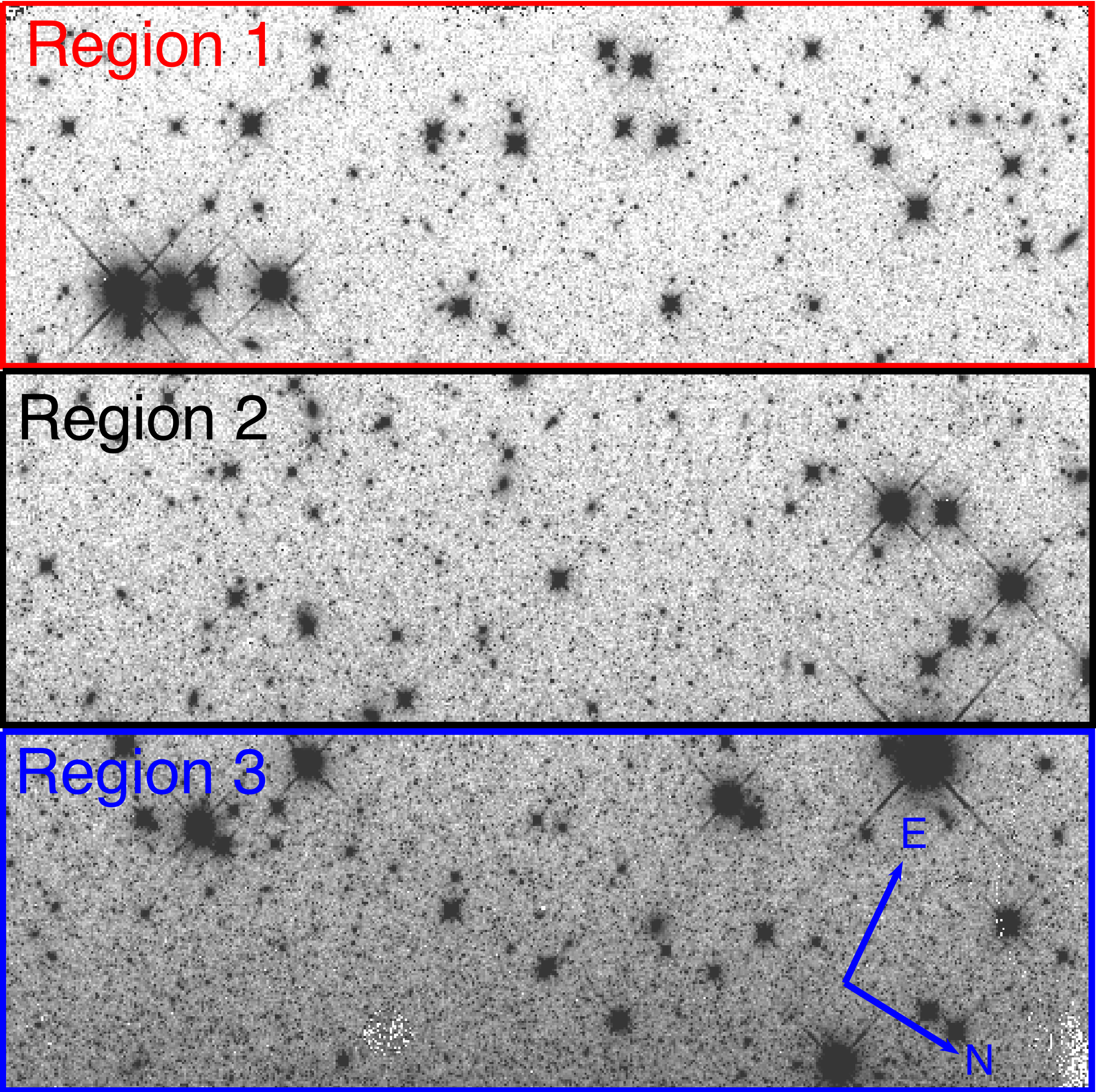}{0.45\textwidth}{(b)}}
\gridline{\fig{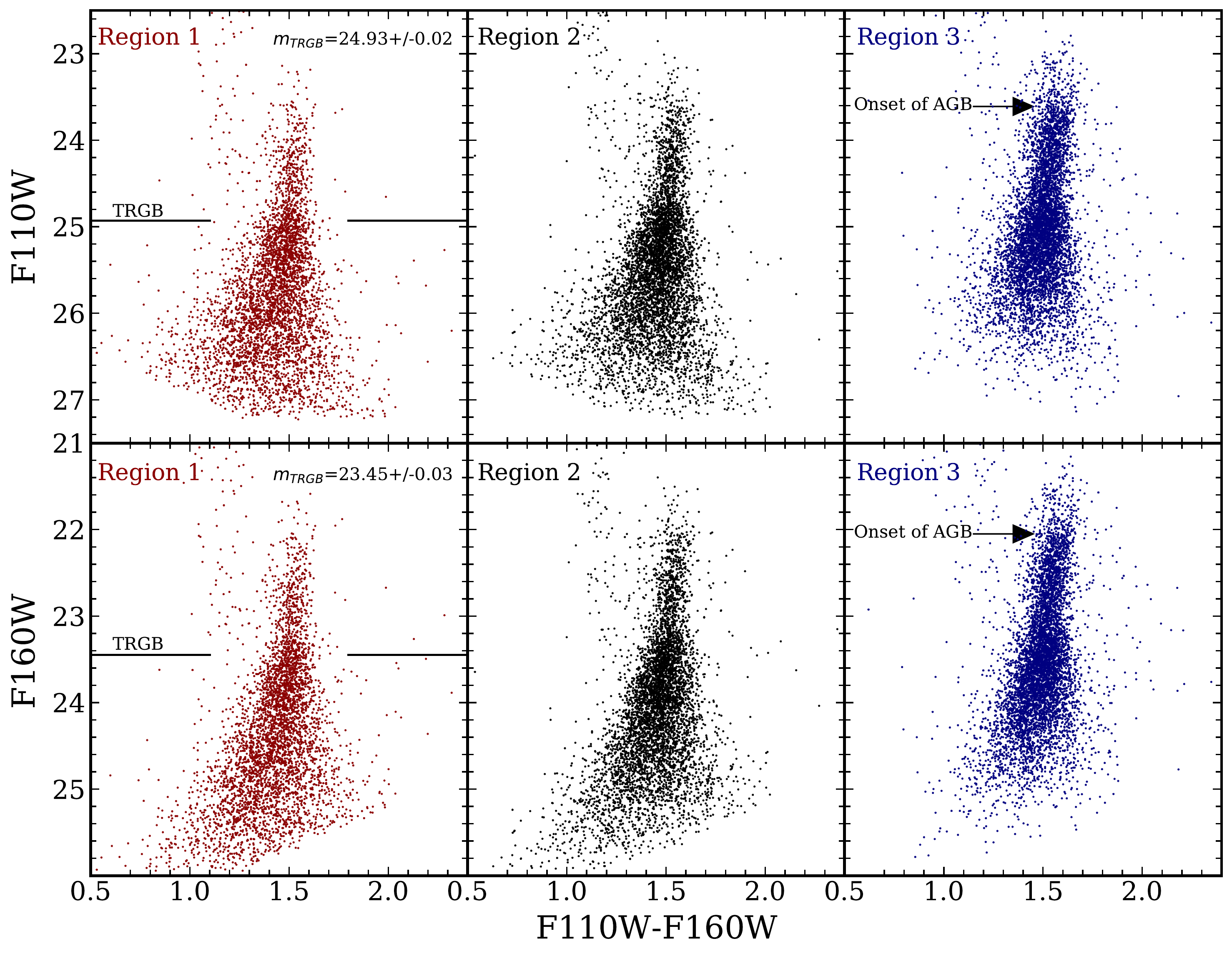}{0.9\textwidth}{(c)}}
\caption{a) Color image of Maffei 2 generated from 2MASS J, H, and $K_{s}$ photometry. The cyan box is the WFC3/IR field of view. b) An F110W exposure: the field is separated into regions based on proximity to the center of Maffei 2. Region 1 is considerably less crowded than Region 3, despite being on the same chip.  c) CMDs generated from each of the three regions in b). Our new distance is calculated from just Region 1. Due to the combined effects of crowding and a large AGB population, the CMD for region 3 is shifted upwards, and the break (incorrectly) assumed to be the TRGB is much brighter in magnitude.\label{maffei2}}
\end{figure*}

\section{Distances}

\subsection{The Tip of the Red Giant Branch}
Stars at the tip of the red giant branch (TRGB) can be used as a standard candle due to their uniform nature just before undergoing the helium flash \citep{dacosta1990,ogTRGB}. The only corrections required are dependencies on metallicity, which are small in the optical \citep{rizzi2007}, and somewhat larger in the near-infrared \citep{NIRbible,wu2014}. TRGB distances to unobscured galaxies within 10 Mpc can be achieved with a single orbit with HST \citep{rizzi2007}, with some programs pushing out to almost 20 Mpc with substantial time investments \citep{CCHP1,CCHP5}.

In brief, our reduction and analysis procedure is as follows. We perform PSF photometry with DOLPHOT \citep{dolphot, dolphot2} on images obtained from the HST archives using the parameters recommended in the user's manual. We then use DOLPHOT to perform artificial star experiments to quantify the levels of photometric errors, bias, and completeness present in the genuine stellar photometry. The stellar catalogs are trimmed to only include sources of the highest quality- for this work, we apply the rigorous ``*.gst'' cuts developed for WFC3/IR photometry by \cite{NIRbible}.

We determine the magnitude of the TRGB with the method described in \cite{MLM}, which involves fitting a broken-power law luminosity function to the AGB and RGB populations, with the break indicating the location of the TRGB. The apparent magnitude and color of the TRGB is corrected for foreground dust extinction, and the metallicity-dependent absolute magnitude of the TRGB is obtained from existing calibrations. This procedure has been used for many papers; for additional recent examples and more in-depth descriptions see \cite{rizzi2017} or \cite{fireworks}.

In the optical, the TRGB is typically determined solely in the F814W filter. In the near-infrared, color magnitude diagrams are developed in both F110W and F160W vs. F110W-F160W. This is because while observations in F110W experience more extinction, the absolute magnitude of the TRGB is only half as sensitive to the metallicity (and hence color) when compared to F160W. Although the analyses are not independent, calculating distances from both filters allows uncertainties to be reduced. 

The DOLPHOT photometry, color-magnitude diagrams, and derived values from this work are all available on the CMDs/TRGB catalog of the Extragalactic Distance Database\footnote{edd.ifa.hawaii.edu} \citep{EDD,EDDCMD}.

\subsection{Maffei 2}
In analyzing their photometry for Maffei 2, \cite{wu2014} found $m_{TRGB}$ = 23.602 $\pm$ 0.037 in the F110W filter, and 22.021 $\pm$ 0.046 in F160W. They also noted the presence of a second break in the luminosity function at F110W = 24.845, but decide that this is unlikely to correspond to the TRGB based on the distances to Maffei 1 and IC 342. 

\cite{tikhonov} performed a reanalysis of this dataset and claim that \cite{wu2014} mistook the tip of the AGB for the TRGB, and that this second discontinuity is the true TRGB. This places Maffei 2 much further than the 3.52 $\pm$ 0.20 Mpc found by \cite{wu2014}, out at 6.83 $\pm$ 0.48 Mpc (though this depends greatly on the assumed color and extinction). 

\cite{tikhonov} also claim the detection of a tip in the parallel ACS observations of Maffei 2 that would correspond with this further distance, though a reanalysis of this data by us shows that the tip feature is deep in the realm of noise. We do not consider their photometric quality cuts to be appropriate; DOLPHOT's `Chi' parameter is not recommended for selections, and there is no mention of any signal to noise or crowding cuts.

To get to the bottom of this confusion, we performed a careful analysis of the stellar populations revealed by the HST photometry. In developing the CMDs to feed into the TRGB fitting software, we first chose to examine the spatial distribution of stars in an interactive fashion with \textit{glue}\footnote{http://docs.glueviz.org/en/stable/index.html}, a Python software package built for astronomical data exploration and visualization. During this process, we found a significant difference in the CMDs obtained from different regions of the field. The difference is quite dramatic (see Figure \ref{maffei2}), more than most cases that we have seen. The regions furthest (Region 1) and closest (Region 3) to the center of Maffei 2 exhibit two main differences. First, the CMD from Region 1 extends $\sim$0.5 magnitudes deeper, as the effects of crowding are substantially less. Second, the discontinuity previously assumed to be the TRGB disappears from Region 3 to Region 1 (radially away from the galaxy), and the fainter discontinuity noted by \cite{wu2014} and \cite{tikhonov} is unambiguously revealed. It is now clear that a significant population of AGB stars have been masking the true TRGB, and that this AGB population is greatly reduced (but not eliminated) by limiting the analysis to the far edge of the WFC3/IR chip.

After limiting the analysis to only Region 1, we perform our TRGB measurements on the two color-magnitude diagrams separately in order to obtain two different (but not completely independent) measurements. We find $m_{TRGB}$ to be 24.93 $\pm$ 0.02 in the F110W filter, and 23.45 $\pm$ 0.03 in F160W. We assume the extinction value of E(B-V) = 1.165 $\pm$ 0.08 from \cite{wu2014}, which were calculated by using the TRILEGAL code \citep{TRILEGAL2009}. TRILEGAL simulates foreground (galactic) stars, and thus allows a robust determination of the extinction value by carefully fitting the simulated galactic population to what is observed in the CMD. This value matches excellently with the value of E(B-V) = $1.17 ^{+0.04}_{-0.06}$ \citep{bayestar17} derived from Pan-STARRS and 2MASS photometry of foreground stars.

We correct the color and magnitude of the TRGB for the derived dust extinction, and apply the absolute magnitude calibrations of \cite{wu2014}.  We find distance moduli of $\mu_{110}$ = 28.81 $\pm$ 0.16, and $\mu_{160}$ = 28.77 $\pm$ 0.18. These results correspond to distances of $D_{110}$ = 5.78 $\pm$ 0.40 Mpc, and $D_{160}$ = 5.67 $\pm$ 0.48 Mpc. We adopt a final distance of d = 5.73 $\pm$ 0.40 Mpc.

\subsection{Maffei 1} 
\begin{figure}
\plotone{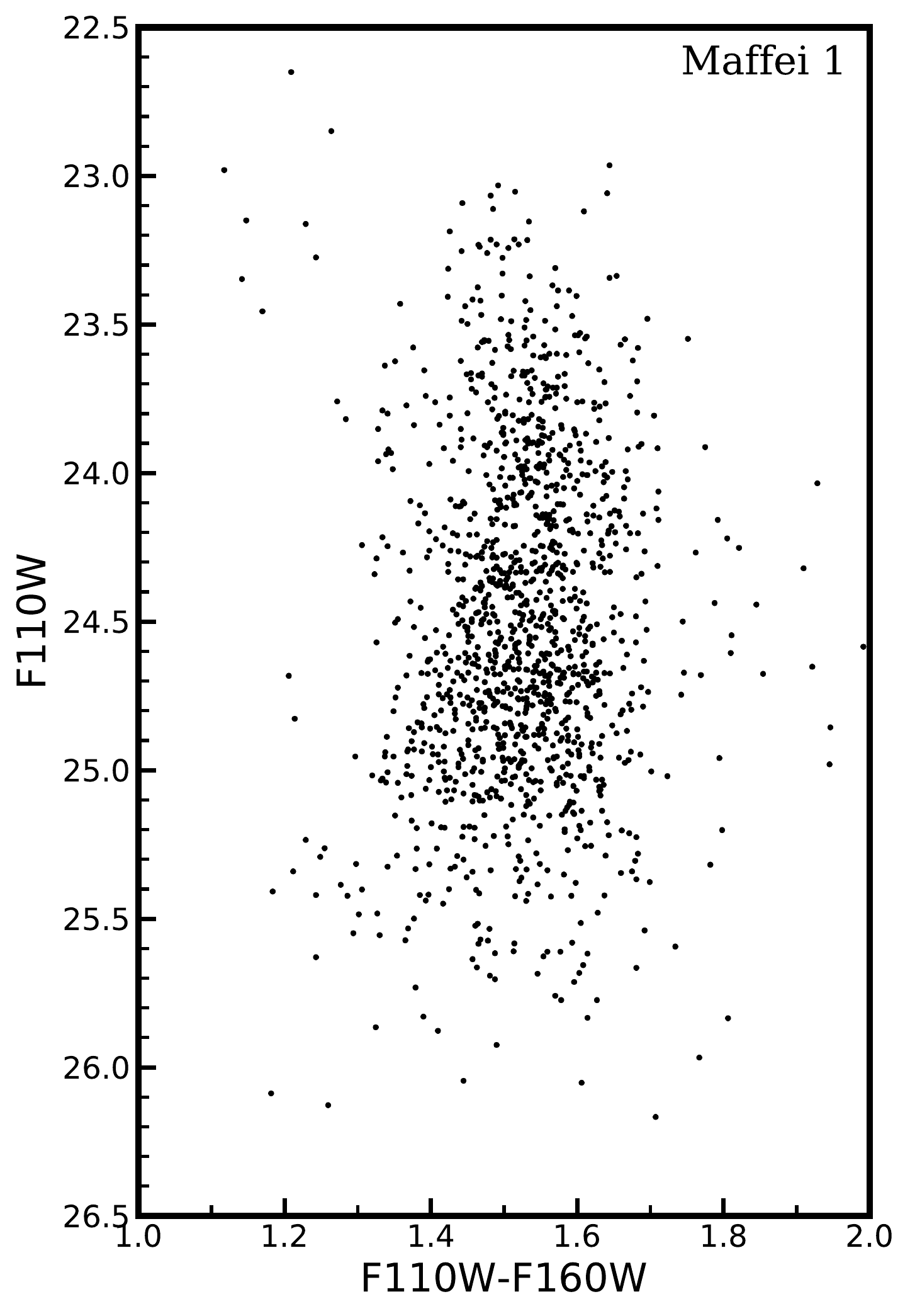}
\caption{A color-magnitude diagram of Maffei 1 obtained from the least crowded region of the WFC3/IR chip. The photometry is $\sim$1 mag less deep than that of Maffei 2 and prohibits a reliable measurement of the magnitude of the TRGB. \label{maffei1CMD}}
\end{figure}

The situation with Maffei 1 is not as clear as with Maffei 2. The issue with crowding is more extreme than with Maffei 2, as the observations peer closer to the center of the galaxy. This results in the final CMDs being $\sim$1 mag less deep when the same photometric cuts are applied. \cite{wu2014} found $m_{TRGB}$ = 23.642 $\pm$ 0.074 in F110W, which likely corresponds to the onset of the AGB in our spatially restricted CMD (see Figure \ref{maffei1CMD}). While there is the hint of a discontinuity present in our CMD at m$\sim$24.5, this is too close to the detection limit to provide a stable result. \cite{tikhonov} quote $m_{TRGB}$ = 24.53 $\pm$ 0.15, but again, we question the reliability of their photometric quality cuts.

Given their similar velocities ($\Delta$$v_{hel}$ = 83~\kms), close proximity (0.67$\degree$), and uncertainty in the distance to Maffei 1, both Maffei 1 and 2 likely lie at the same distance, and so we place Maffei 1 in the same group as Maffei 2. 
 
\subsection{IC 342}
The distance to IC 342 is much more well-constrained, as there is less than half the foreground extinction when compared to Maffei 1 or Maffei 2. This allowed \cite{wu2014} to find distances with both the ACS and WFC3/IR observations. Their final distance of 3.45 $\pm$ 0.13 Mpc also matches closely with the value from \cite{tikhonov} of 4.02 $\pm$ 0.30 Mpc, with the difference mainly due to differences in adopted extinction coefficients. 

Since the data obtained from the HST archives receives regular recalibration, while DOLPHOT also receives regular enhancements, we decided to redo the analysis. First, we do not find the large spatial variation in CMDs seen with Maffei 1 and Maffei 2. We obtain distance moduli of $\mu_{110}$ = 27.56 $\pm$ 0.10, and $\mu_{160}$ = 27.57 $\pm$ 0.13, which correspond to distances of $D_{110}$ = 3.25 $\pm$ 0.15 Mpc, and $D_{160}$ = 3.27 $\pm$ 0.19 Mpc. These closely match the near-infrared distances ($D_{110}$ = 3.31 $\pm$ 0.20 Mpc, and $D_{160}$ = 3.30 $\pm$ 0.22 Mpc) of \cite{wu2014}.

\section{The Maffei Group \& Peculiar Velocity}

\cite{kourkchi2017} assign 8 galaxies to the Maffei Group.  Figure~7 in \cite{wu2014} shows the projected distribution of these galaxies, illustrating the proximity on the sky of the IC~342 Group and the seriousness of obscuration in the Maffei direction.  Here we would add two galaxies to the Maffei Group, KKH5 with $V_{helio} = 61$~\kms\ and KKH12 with $V_{helio} = 70$~\kms, but reject one galaxy, KKH11, distant in projection from the group center and with the discordant velocity $V_{helio} = 296$~\kms.

Of the 9 galaxies considered as group members, only Maffei~2 has a reliable distance.  Table~\ref{members} identifies the members and their observed velocities.  It is seen that Maffei~2 has the most negative velocity, although not statistically discrepant.  On the other hand, the dwarf galaxy MB1 has a velocity substantially above the group mean.  MB1 is only 30~kpc in projection from the dominant galaxy Maffei~1.  The group mean, absent MB1, is $302 \pm13$~\kms\ in the Local Sheet reference frame \citep{peculiar} with an r.m.s. dispersion 38~\kms.  The velocity of MB1 of 430~\kms\ is 128~\kms\ higher which would not be unusual for a near satellite of a massive galaxy.  Including MB1 in an unweighted averaging, the group velocity is $316 \pm18$~\kms\ with dispersion 55~\kms. 

Accepting the one reliable distance of Maffei~2 for the group of $d = 5.73\pm0.40$~Mpc, the anticipated Hubble expansion velocity $H_0 d = 430\pm30$~\kms\ assuming $H_0 = 75$~\kmsMpc, consistent with the $Cosmicflows-3$ compendium of distances \citep{CF3}. The dependence on the assumed value of $H_0$ is weak; if the value 73 is assumed then the expansion velocity is 420~\kms.  Continuing with the assumption of $H_0 = 75$~\kmsMpc, the Maffei Group has a radial peculiar velocity of $-128\pm33$~\kms\ if we accept the group velocity without MB1 of 302~\kms\, or $-114\pm35$~\kms\ if MB1 is included. In either case and any reasonable value of $H_0$, the Maffei Group has a significant peculiar velocity toward us. 

\subsection{Relation to the IC~342 Group}

We confirm the previous measurements of the distance of IC~342, here deriving $3.26\pm0.15$~Mpc with our near infrared measurement.  In combination with the optical measurement, we accept a distance of $3.45\pm0.13$~Mpc \citep{wu2014}.  \cite{kourkchi2017} identify 8 galaxies associated with the IC~342 Group, 7 with TRGB distances.  The weighted average of the moduli gives a group distance of $3.22\pm0.10$~Mpc. The averaged group velocity is $-36$~\kms\ heliocentric and $180\pm27$~\kms\ in the Local Sheet frame, with dispersion 72~\kms.  The anticipated Hubble expansion velocity for the group is $242\pm8$~\kms\ so the radial peculiar velocity is $-60\pm28$~\kms. 

With the nearer distance for the Maffei Group given by \cite{wu2014}, IC~342 would only be 700~kpc from Maffei 1 and 2 and all the galaxies associated with these principal galaxies would be within a common infall region. However at the greater distance advocated here, the Maffei Group is $2.5\pm0.5$~Mpc separated from the IC~342 Group and these two entities are quite distinct.

\begin{table}
\begin{tabular}{|l|c|c|c|c|c|}
\hline
\textbf{Galaxy Name} & \textbf{Ty} & \textbf{$K_{s}$} & \textbf{Log($L_{K}$)} & \textbf{$V_{h}$} & \textbf{$V_{LS}$} \\ \hline
Maffei 1        & E           & 4.28        & 11.12           & 66            & 305            \\ \hline
Maffei 2        & Sbc         & 4.93        & 10.85           & -17           & 220            \\ \hline
Dwingeloo 1     & Sbc         & 8.37        & 9.48            & 112           & 341            \\ \hline
Dwingeloo 2     & Irr         & 10.2        & 8.75            & 94            & 325            \\ \hline
MB1             & Sdm         & 10.5        & 8.63            & 190           & 430            \\ \hline
KK22            & Irr         & 10.9        & 8.47            & 59            & 289            \\ \hline
KKH6            & Irr         & 11.8        & 8.11            & 53            & 294            \\ \hline
KKH5            & Irr         & 13.5        & 7.43            & 61            & 294            \\ \hline
KKH12           & Irr         & 13.5        & 7.43            & 70            & 311            \\ \hline
\end{tabular}
\caption{A summary of the members of the Maffei group, including galaxy type, reddening corrected $K_{s}$ magnitudes, $L_{K}$ luminosity ($L_{\odot}$) assuming $d = 5.73$~Mpc, and velocities in the heliocentric and Local Sheet reference frames (\kms).}

\label{members}
\end{table}

\section{Summary}
\cite{tikhonov} were correct in questioning the distance to the Maffei Group found in our earlier work.  The AGB and RGB can be easily confused, especially at near infrared bands where the two stellar branches have similar colors.  The situation with the Maffei Group is aggravated by substantial obscuration and crowding. By limiting our analysis to the region furthest from the center of Maffei 2, we are able to clearly isolate the true TRGB. In detail, we find a distance to Maffei~2 of $5.73\pm0.40$~Mpc, significantly closer than the value of $6.83\pm0.48$~Mpc given by \cite{tikhonov}.  We cannot offer a distance to Maffei~1 with the observational material available.

The Maffei and IC~342 groups should be seen as two distinct entities separated by 2.5~Mpc.  Both groups have peculiar velocities toward us, of $-134$ and $-60$~\kms\ respectively.  Both entities lie with us on the supergalactic equator so we cannot see components of motion perpendicular to the equatorial sheet, as has become familiarly associated with evacuation of the Local Void \citep{rizzi2017, shaya2017}.  Nonetheless, the mapping of flows with the full $Cosmicflows-3$ compilation of distances anticipates flows toward us from the direction of the Maffei Group.  It is well known that there is a major void to the foreground of the Perseus-Pisces complex at 5,000~\kms\ \citep{hg1986}.  Indeed, it is becoming evident that the Local Void and the void in front of the Perseus-Pisces filament are one and the same.  The Maffei and IC~342 groups are projected in front of the Perseus-Pisces structure and the void in that direction.  We are witnessing expansion of the void toward us.  

We are currently analyzing distances to several other galaxies on the fringes of the Local Void and will reserve further discussion on the structure and dynamics of the Local Void to a later paper.

\section{Acknowledgements}
G.A. would like to thank Connor Auge and Michael Tucker for useful discussions throughout the length of the paper. 

All of the data presented in this paper were obtained from the Mikulski Archive for Space Telescopes (MAST). STScI is operated by the Association of Universities for Research in Astronomy, Inc., under NASA contract NAS5-26555. Support for this work was provided by NASA through grant number HST-AR-14319 from the Space Telescope Science Institute. I.K. acknowledges support by RFBR grant 18-02-00005.

This research made use of Montage, funded by the National Aeronautics and Space Administration's Earth Science Technology Office, Computational Technnologies Project, under Cooperative Agreement Number NCC5-626 between NASA and the California Institute of Technology. The code is maintained by the NASA/IPAC Infrared Science Archive.



\end{document}